\documentclass[aps,pra,twocolumn,showpacs,groupedaddress]{revtex4} 

\usepackage{graphicx}  
\usepackage{dcolumn}   
\usepackage{bm}       
\usepackage{amssymb}   
\usepackage{amsmath}

\begin{document}

\title{Quantum projection noise limited interferometry with coherent atoms in a Ramsey type setup}

\author{D.\,\,D\"oring$^{1,2}$}
\author{G.\,\,McDonald$^{1,2}$}
\author{J.\,E.\,\,Debs$^{1,2}$}
\author{C.\,\,Figl$^{1,2}$}
\author{P.\,A.\,Altin$^{1,2}$}
\author{H.-A.\,\,Bachor$^{1,2}$}
\author{N.\,P.\,\,Robins$^{1,2}$}
\author{J.\,D.\,\,Close$^{1,2}$}

\affiliation{$^1$Australian Research Council Centre of Excellence for Quantum-Atom Optics, The Australian National University, Canberra, 0200, Australia}

\affiliation{$^2$Department of Quantum Science, Research School of Physics and Engineering, The Australian National University, Canberra, 0200, Australia}

\date{\today}

\begin{abstract}
Every measurement of the population in an uncorrelated ensemble of two-level systems is limited by what is known as the quantum projection noise limit. Here, we present quantum projection noise limited performance of a Ramsey type interferometer using freely propagating coherent atoms. The experimental setup is based on an electro-optic modulator in an inherently stable Sagnac interferometer, optically coupling the two interfering atomic states via a two-photon Raman transition. Going beyond the quantum projection noise limit requires the use of reduced quantum uncertainty (squeezed) states. The experiment described demonstrates atom interferometry at the fundamental noise level and allows the observation of possible squeezing effects in an atom laser, potentially leading to improved sensitivity in atom interferometers.    

\end{abstract}

\maketitle 

\section{Introduction}

Since the first realization of Bose-Einstein condensates (BECs) in 1995 \cite{Anderson1995,Davis1995,Bradley1995}, the focus of experiments with coherent atomic samples has shifted from the investigation of fundamental properties of BECs towards the application of these properties to probing a variety of physical processes. Bose-Einstein condensates have been extensively used to simulate lattice effects in solid state physics \cite{Foelling2006,Spielman2007,Bloch2008}, to investigate dipolar quantum gases \cite{Lahaye2007} and to show Anderson localization of matter waves \cite{Billy2008,Roati2008}, to mention only a few. Regarding atom interferometry \cite{Cronin2007}, however, the most remarkable results have been so far achieved using thermal atoms, e. g. for the measurement of the gravitational constant $G$ \cite{J2007,G2008}, the fine structure constant $\alpha$ \cite{S2002,Malo2008} and for the definition of time in atomic clocks \cite{K2007,Ch2002}. Bose-Einstein condensates and the coherent beams or pulses derived from them, known as atom lasers, offer a compelling alternative to thermal sources. Atom lasers like the $^{87}$Rb device used in this experiment are narrow-linewidth, highly directed beams, and their narrow transverse velocity widths make them favorable for high momentum transfer beam splitting in atom interferometers, effectively allowing larger enclosed areas and higher sensitivities to, e.g., accelerations and rotations. The internal atomic states of the freely propagating atom laser used in this experiment ($\left|F=1,m_F=0\right\rangle$ and $\left|F=2,m_F=0\right\rangle$) are magnetically insensitive to first order. Similar to atoms in an atomic clock, the atom laser is strongly decoupled from its environment and well isolated from external sources of noise.

Improving the applicability of BECs and atom lasers to the field of atom interferometry will involve either increasing the average atomic flux or improving the interferometric sensitivity by implementing quantum mechanical reduced-uncertainty (so-called squeezed) states \cite{ueda}, allowing a decrease in detection noise below what is known as the quantum projection noise limit \cite{W1993,D1994}. Squeezing of atomic samples has become a subject of increasing importance in recent years. Est$\grave{\mbox{e}}$ve {\it et al.} \cite{JE2008} have demonstrated spin squeezing in trapped Bose-condensed samples, and groups in Copenhagen and Cambridge have observed spin squeezing on the atomic clock transition in $^{133}$Cs \cite{J2008} and $^{87}$Rb \cite{Monika2008}. Atom lasers, coherent atomic beams output-coupled from a Bose-Einstein condensate, offer a promising choice to implement a variety of squeezing mechanisms \cite{Mattias2007,Simon2009}. A necessary prerequisite for implementing and observing spin-squeezing in an atomic sample is a highly stable interferometry and detection setup. More precisely, one needs to operate at the quantum projection noise limit.

In this paper, we present quantum projection noise limited operation of our atom laser Ramsey type interferometer \cite{D2009}. We use a novel optical setup based on a Sagnac interferometer to achieve highly passively stable two-photon Raman coupling between the internal atomic states involved in the interferometer. The stability of the interferometer using this coupling mechanism is analyzed, using either a single $\frac{\pi}{2}$-beamsplitter or a complete $\frac{\pi}{2}$-$\frac{\pi}{2}$-Ramsey sequence \cite{Norman1950}. We measure the probability $p=N_2/(N_1+N_2)$ of transferring atoms between two internal states, $\left|1\right\rangle$ and $\left|2\right\rangle$, using a freely propagating pulse of coherent atoms. In the case of equal atom numbers ($N_1=N_2$, $p=0.5$), fluctuations in the absolute number $N=N_1+N_2$ do not contribute to the classically achievable minimum uncertainty in $p$. The limit is determined by quantum projection noise, $\sigma_p=1/\sqrt{4N}$. A comparison between this expected uncertainty and the experimentally measured noise shows that quantum projection noise is the limiting factor in our apparatus. The best achieved detection uncertainty lies below $0.5\,\%$, corresponding to the quantum projection noise limit of a sample of $N= 10^4$ atoms.

The paper is organized as follows. In section \ref{scheme}, we describe the experimental scheme, in particular the inherently stable optical Sagnac interferometer used to produce Raman light and the atomic shot noise limited imaging system. Section \ref{results} contains the results of our Ramsey interferometry experiment and a noise analysis in the context of the quantum projection noise limit. We conclude the paper with section \ref{conclusion}.

\section{Experimental scheme}
\label{scheme}

The reduction of noise to an absolute minimum is the critical point that every part of the experimental setup is focussed on. In the following, we describe the scheme for producing Raman light to couple two internal states of the freely propagating atom laser. This stable coupling allows us to observe quantum projection noise on a Ramsey-type interferometric signal. The signal is detected via shot-noise limited absorption imaging, as discussed in the last part of this section. 

\subsection{Sagnac interferometer setup}
\label{Sagnac}

\begin{figure}[t]
\includegraphics[scale=1.35]{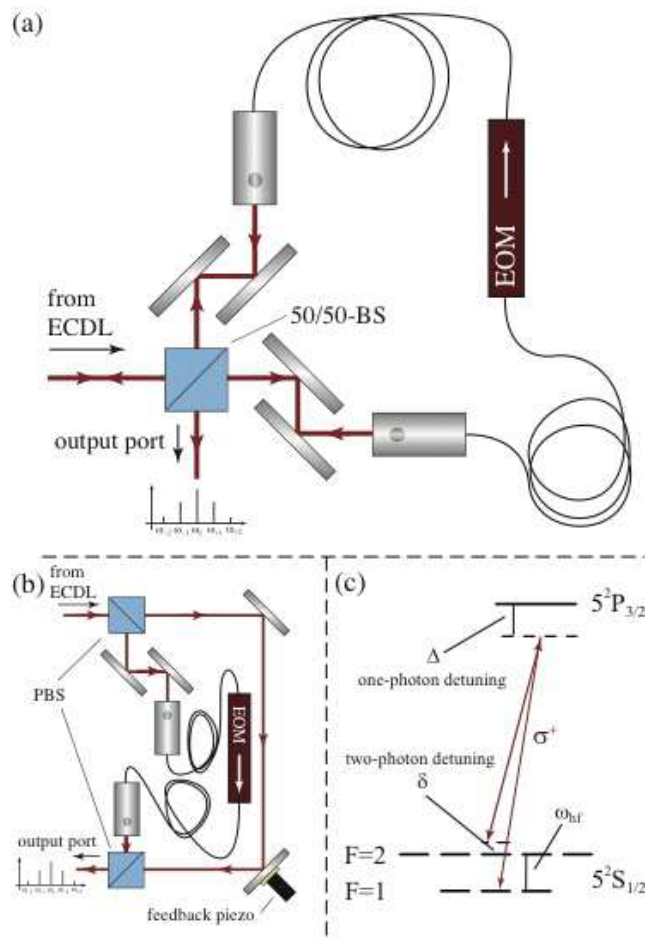}
\caption{\label{setup}(Color online) (a) Sagnac interferometer setup, based on a non-polarizing $50/50$-beamsplitter. The operation of the interferomter relies on the directivity of the electro-optic phase modulator. (b) Equivalent setup using a Mach-Zehnder type design. (c) Level scheme for Raman coupling of the states $\left|1\right\rangle$ ($F=1,m_F=0$) and $\left|2\right\rangle$ ($F=2,m_F=0$).}
\end{figure}

The internal state coupling scheme is crucial for achieving a stable interferometric signal. We implement a two-photon Raman scheme to couple the states $\left|1\right\rangle \equiv \left|F=1,m_F=0\right\rangle$ and $\left|2\right\rangle \equiv \left|F=2,m_F=0\right\rangle$ of $^{87}$Rb (Fig. \ref{setup}(c)). As opposed to the direct application of microwaves, Raman coupling allows for a straightforward implementation of spatially selective coupling and is extendable to external state beamsplitting. A fiber-coupled electro-optic modulator (EOM) is used to generate sidebands on the light from an external cavity diode laser (ECDL). The phase modulation imprinted onto the electric field by the EOM can be written as
\begin{equation}
{\bf E}_\text{pm}(t) = {\bf E}_0e^{i\left(\omega t + \frac{\phi}{2} \cos \left(\omega_\phi t\right)\right)} + c.c.,
\end{equation}
where $\omega$ is the optical frequency of the laser, $\phi/2$ the amplitude and $\omega_\phi$ the frequency of the phase modulation. Decomposing the phase modulation into sidebands gives
\begin{equation}
e^{i \frac{\phi}{2} \cos\left(\omega_\phi t\right)}=\sum_{n=-\infty}^{\infty}i^n J_n\left(\frac{\phi}{2}\right)e^{i n\omega_\phi t}\,.
\end{equation}
The amplitude of each sideband is proportional to the n$^\text{th}$ order Bessel function of the first kind, $J_n\left(\phi/2\right)$. 

In the limit of large detuning, the semiclassically calculated Rabi frequency for two-photon Raman coupling between two states $\left|1\right\rangle$ and $\left|2\right\rangle$ is $\Omega_\text{R}=\Omega_a \Omega_b/(4\Delta)$, where $\Delta$ is the detuning from one-photon resonance (see Fig. \ref{setup}(c)), and $\Omega_a$ and $\Omega_b$ are the one-photon Rabi frequencies for the two light fields, defined by $\Omega_{(a,b)}=\frac{{\bm \mu}_{ge}\cdot{\bf E}_{0,(a,b)}}{\hbar}$, where ${\bm \mu}_{ge}$ is the dipole matrix element quantifying the coupling strength between the ground and the excited state.

The most straightforward way to drive Raman transitions with the phase modulated field would seemingly be using the pure modulated light, equivalent to setting ${\bf E}_a={\bf E}_b={\bf E}_\text{pm}$. For large $\Delta$ (significantly larger than the excited state hyperfine splitting), however, it is not possible to drive Raman transitions in an atomic sample that way (see Refs. \cite{JD2008,P2003}). Driving Raman transitions on two-photon resonance (corresponding to $\delta=0$ in Fig. \ref{setup}(c)) requires the atomic transition frequency $\omega_{\text{hf}}$ to be a multiple of the phase modulation frequency $\omega_\phi$. The resulting two-photon Rabi frequency $\Omega_\text{R}$ is proportional to $\sum_{n=-\infty}^{\infty}\sum_{m=-\infty}^{\infty}i^{n-m}J_n\left(\phi/2\right)J_m\left(\phi/2\right)^*e^{i(n-m)\omega_\phi t}+c.c.$. The terms in this sum that contribute to the overall two-photon Rabi frequency interfere destructively, leaving  $\Omega_\text{R}=0$. To overcome this destructive interference, the signal from the EOM must be at least partly converted into an amplitude modulated electric field. There are a variety of ways to achieve this goal, such as using cavities to strip off sidebands and produce separate beams containing only one optical frequency. Alternatively, one can interfere the phase modulated light with light at the carrier frequency, thereby removing the destructive interference in $\Omega_\text{R}$ and allowing the Raman transitions to be driven with one beam containing all frequencies involved. Ideally, one wants to implement such a mechanism without introducing additional noise onto the coupling.

\begin{figure}[t]
\includegraphics[scale=0.75]{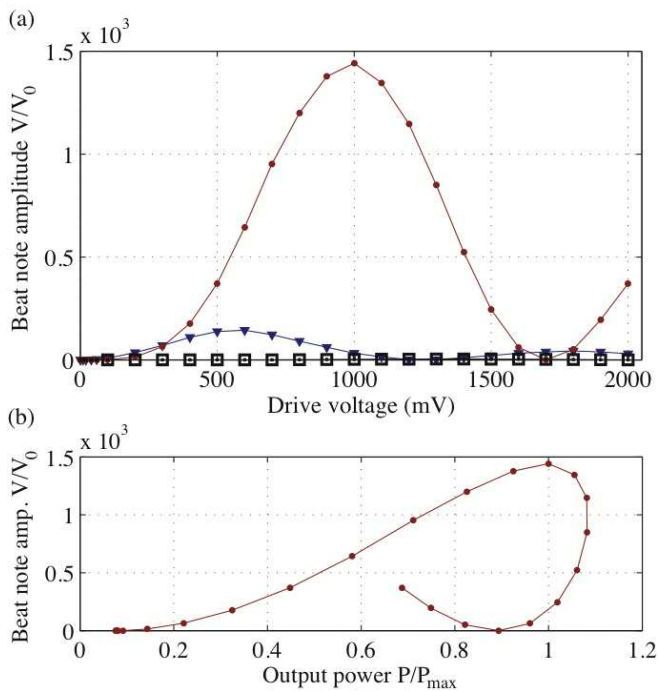}
\caption{\label{beat note}(Color online) (a) Measured beat note amplitude as a function of drive voltage for a microwave frequency of $1\,$GHz. $\text{V}_\text{0}$ is an arbitrary reference voltage. The beat note is measured at $1\,$GHz (blue triangles), $2\,$GHz (red dots) and $3\,$GHz (black squares). (b) Calibration curve for the beat note amplitude as a function of output power of the Sagnac interferometer. The frequencies of the beat note and phase modulation are $2\,$GHz and $1\,$GHz, respectively. The power is normalized to the power at maximum beat note amplitude.}
\end{figure}

In previous work \cite{D2009}, an actively stabilized Mach-Zehnder interferometer (Fig. \ref{setup}(b)) was used, with the fiber-coupled EOM placed in one arm of the setup. Stabilization of this setup was difficult, mainly due to large temperature related phase drifts in the fiber, leading to fluctuations in the interferometer pathlength difference of multiple wavelengths and a strongly fluctuating two-photon Rabi frequency $\Omega_\text{R}$ experienced by the atoms. The instabilities in the Mach-Zehnder interferometer were the limiting source of noise on the previous atom laser interferometry experiment \cite{D2009}. Compared to that, the setup used in the experiment described here has been significantly improved by using a Sagnac interferometer to produce an amplitude modulated light field. We take advantage of the directivity of the EOM device and place it in the Sagnac setup (Fig. \ref{setup}(a)), thereby interfering the phase modulated light with an unmodulated field of equal amplitude. The electric field at the output port can be writen as

\begin{figure}[t]
\includegraphics[scale=0.8]{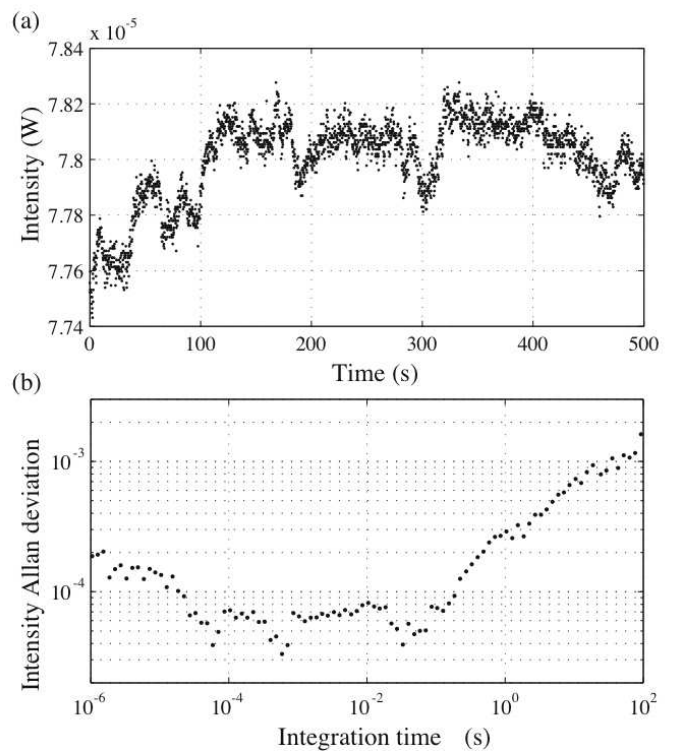}
\caption{\label{allan}(a) Measured fluctuations of the summed intensity of both beamsplitters (for a drive voltage of $1.5\,$V at $3.417\,$GHz) and (b) Allan deviation of the relative intensity measured at the same position. For long integration times, polarization fluctuations in the EOM-fiber and the transfer fiber dominate the measured Allan deviation.}
\end{figure}

\begin{eqnarray*}
{\bf E}(t)={\bf E}_0(t)e^{i\omega t}\left[ae^{i(\pi+\delta\phi_0)}+\sum_{n=-\infty}^{\infty}i^n J_n\left(\frac{\phi}{2}\right)e^{i n\omega_\phi t}\right]\\
+ \textit{c.c.}\,,
\end{eqnarray*}
where $\delta\phi_0$ accounts for any remaining phase difference (e.g. due to misaligned polarizations in the fiber) between the two directions in the Sagnac interferometer. For equal field amplitudes in both directions ($a=1$) and no phase modulation on the EOM, the intensity of the Sagnac output port vanishes due to destructive interference of the light fields propagating in opposite directions. Increasing the modulation depth leads to an increase in intensity at the output port and an increase in the two-photon Rabi frequency $\Omega_\text{R}$. In the limit of large detuning $\Delta\rightarrow\infty$, $\Omega_\text{R}$ is proportional to the intensity beat signal.

The resonance condition for hyperfine transitions to be driven in the atomic sample (using the setup described above) requires the transition frequency to be a multiple of the modulation frequency, $\omega_\text{hf}=m\omega_\phi$. Two inherently distinct cases have to be considered here. For odd values of $m$ ($1,3,5,...$), the two-photon Rabi frequency $\Omega_\text{R}$ is suppressed due to destructive interference of the different terms contributing to $\Omega_\text{R}$. This suppression is based on the same mechanism that prevents Raman transitions to be driven with a purely phase-modulated light field. In the case of even values of $m$ ($2,4,6,...$), the situation is different, and the electric field at the output port of the Sagnac interferometer is suitable for driving Raman transitions in the atomic sample. Its amplitude modulation is no longer suppressed by the destructive interference effect described above.

To verify these points, we measure the beat note amplitude on the light at the output of the Sagnac interferometer for various modulation depths at a modulation frequency of $1\,$GHz and values of $m=1,2$ and $3$ (Fig. \ref{beat note}(a)). The maximum beat note amplitude is strongest for $m=2$ (red dots in Fig. \ref{beat note}(a)), and is suppressed for the cases $m=1$ (blue triangles) and $m=3$ (black squares). The measurement is in excellent qualitative agreement with our basic theoretical picture. In the experiment described here, we use $m=2$ and drive the EOM at $3.417\,$GHz --- half the transition frequency between the states $\left|1\right\rangle$ and $\left|2\right\rangle$.

A useful feature of our setup for generating the appropriate light field is that both the output intensity and the beat note amplitude after the Sagnac interferometer have a well defined dependency on the modulation depth of the phase modulator. For a given output intensity, we can therefore infer the beat note amplitude on the light field. This relation of output intensity and beat note amplitude is independent of the modulation frequency. In other words, simultaneously measuring the intensity output and the beat note amplitude after the Sagnac interferometer allows us to calibrate the beat note amplitude, and thus $\Omega_\text{R}$, with respect to the intensity. As the calibration is independent of the modulation frequency of the EOM, it can be carried out at any arbitrary frequency within the bandwidth of the EOM. Once such a calibration curve has been established, one can infer the beat note amplitude from a simple intensity measurement of the light field. For high microwave frequencies, this is a significant advantage compared to a direct beat note measurement which requires high bandwidth detection devices. Figure \ref{beat note}(b) shows the measured calibration curve for our setup. To obtain high microwave frequency beat note amplitudes, it is thus sufficient to measure the Sagnac interferometer intensity output and derive the beat note amplitude from the calibration depicted in Fig. \ref{beat note}(b).

\begin{figure}[t]
\includegraphics[scale=0.6]{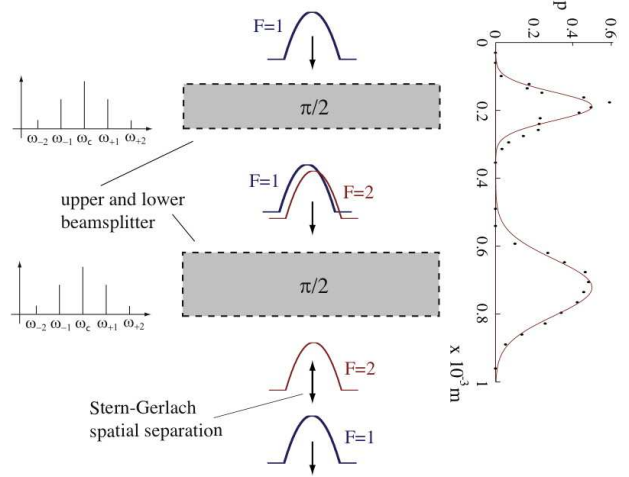}
\caption{\label{Ramsey_setup} (Color online) Schematic drawing of the interferometer setup. An atom laser pulse originally in state $\left|1\right\rangle$ travels through two spatially separated beamsplitters, each containing frequencies with a spacing of $\omega_\text{hf}/2$. The plot on the right depicts a measurement of the spatial width of each of the light sheets.}
\end{figure}

We use a high-accuracy microwave generator (Rhode\&Schwarz SMR20) stabilized to an external frequency reference (Stanford Research Systems FS725) to drive the EOM, leading to a highly stable beat frequency of the Sagnac interferometer output. Fluctuations in the Sagnac setup are common mode to the two equally polarized beams travelling in opposite directions in the interferometer. Due to the inherent passive stability of the system, we are able to avoid active stabilization loops. Figure \ref{allan} illustrates the intensity stability of the Sagnac interferometer output measured at a position equivalent to the position of the atomic sample. The maximum Allan deviation \cite{Allan} of the relative intensity corresponds to an uncertainty in the interferometer transition probability $p$ of $\sigma_p=2\times 10^{-3}$, below the quantum projection noise limit for the atom numbers used in this experiment (see section \ref{results}). 

\begin{figure}[t]
\includegraphics[scale=0.92]{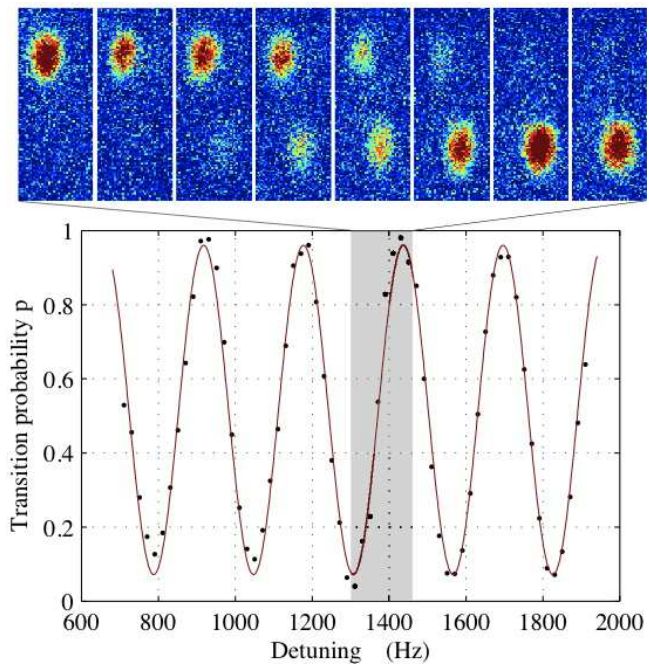}
\caption{\label{fig:fringes} (Color online) Central part of a typical Ramsey fringe measurement. The atom cloud pictures on the top show shot noise limited absorption images.}
\end{figure}

\subsection{Ramsey interferometry with freely propagating coherent atoms}

Having produced a light field capable of driving Raman transitions in our atomic sample, we couple the light through a polarization maintaining optical fibre to the position of our atomic samples. Via a birefringent calcite crystal and two cylindrical lenses ($f=400\,$mm and $f=70\,$mm), we realize two spatially separated light sheets. These serve as beamsplitters for the atom interferometer, as described in \cite{D2009}. Details of our experimental setup for making Bose-Einstein condensates can be found in \cite{D2008}. In brief, we produce $^{87}$Rb Bose-Einstein condensates in the $\left|F=1,m_F=-1\right\rangle$ state, trapped in a quadrupole-Ioffe configuration magnetic trap and rf output-couple an atom laser in the magnetically insensitive $\left|F=1,m_F=0\right\rangle$ state. The atom laser propagates freely under gravity and travels through the two beamsplitters where it is coupled to the $\left|F=2,m_F=0\right\rangle$ state (see Fig. \ref{Ramsey_setup}). The Raman coupling is implemented using copropagating beams, resulting in negligible momentum transfer in each beamsplitter.

\subsection{Imaging system}

The detection system is a critical piece of the setup. Atomic shot noise limited operation is required in order to be able to reduce the detection noise level below the quantum projection noise limit. We use a highly stable detection setup, including absorption imaging onto a CCD camera of the type PointGrey GRAS-14S5M/C. All classical noise sources are reduced to an absolute minimum. It is of particular importance to remove all interference effects on the absorption images. The two states involved in the interferometer are separately detected via a Stern-Gerlach type spatial separation, relying on the opposite second order Zeeman shift of the states $\left|1\right\rangle$ and $\left|2\right\rangle$. We apply a pulsed magnetic field gradient for a time long enough to clearly separate the positions of the two states. Figure \ref{fig:fringes} contains a sequence of atom clouds, detected with the described imaging system.

\begin{figure}[t]
\includegraphics[scale=0.98]{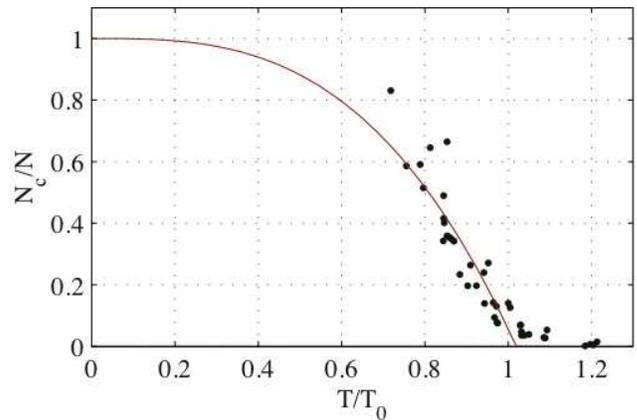}
\caption{\label{fig:number_calibration} (Color online) Atom number calibration curve. The graph shows the condensate fraction as a function of normalized temperature of the atomic sample. The red line is a fit of the form $N_c/N=1-(T/a T_0)^3$, giving an intercept of $a=1.019\pm0.007$.}
\end{figure}

The atom number is measured pixelwise via the standard absorption imaging procedure, using 
\begin{equation}
\label{eq:imaging}
N_\text{px}=c_0 \left(L \ln\frac{I_0}{I} + \frac{I_0-I}{I_\text{sat}}\right),
\end{equation}
where $c_0=2\pi (px)^2/(3 \lambda^2 M^2)$ contains the resonant cross section $2\pi/(3 \lambda^2)$, the pixel area $(px)^2$ and the imaging magnification $M$. A possible detuning $\Delta_\text{im}$ of the imaging light is accounted for with $L=(4 \Delta_\text{im}^2 + \Gamma^2)/\Gamma^2$, where $\Gamma$ is the natural linewidth of the imaging transition $\left|F=2,m_F=2\right\rangle\rightarrow\left|F'=3,m_{F'}=3\right\rangle$ in $^{87}$Rb. $I$ is the intensity in the atom picture and $I_0$ the intensity in an identical picture containing no atoms. The saturation intensity on the imaging transition has been measured and is given by $I_\text{sat}$. We calculate the total atom number in a specific internal atomic state by summing over the pixels in a well-defined region of the CCD chip containing only this one atomic state. 

The detection system has been calibrated to ensure accurate atom number counting. We use the well known dependence of the critical condensation temperature $T_0$ on the number of atoms $N$ in the atomic sample. For finite atom numbers \cite{Wolfgang1996},
\begin{equation}
T_0=\frac{\hbar\overline{\omega}}{k_B}\left(\sqrt[3]{\frac{N}{\zeta(3)}}-\frac{\pi^2}{12 \zeta(3)}\right),
\end{equation}
where $\overline{\omega}=(\omega_x \omega_y \omega_z)^{1/3}$ is the geometric mean of the trapping frequencies and $\zeta(s)$ the Riemann zeta function. A fit of the form $N_c/N=1-(T/a T_0)^3$, where $a$ is the fitting parameter and $N_c$ the number of atoms in the condensate fraction of the cloud, yields $a=1.019\pm0.007$, in reasonable agreement with the expected value of $a=1$ (see Fig. \ref{fig:number_calibration}). The small deviation of the measured value is probably due to imperfect polarization of the imaging light, causing undercounting of the atom number and therefore a value of $a>1$. In addition, the fitting routine used to determine the number of atoms in the condensate and the thermal fraction introduces a small systematic error. At temperatures slightly above the condensation temperature, the thermal cloud exhibits a Bose-enhanced density distribution, which can be wrongly interpreted as a superposition of a thermal cloud and a small Bose-Einstein condensate. This effect will lead to an apparently less sharp phase transition, causing a value of $a>1$. The overall deviation of $a$ from $1$ translates into a systematic atom number counting error of $5.8\,\%$.

The accuracy of the above described atom number calibration relies critically on a correct temperature measurement. We use time-of-flight absorption images taken $t_\text{exp}=24.1\,$ms after releasing an atom cloud from the trap and measure the width of the thermal fraction of the expanded cloud. A two-dimensional bimodal fit is performed, taking into account both the Thomas-Fermi profile of the condensed fraction of the cloud and the Gaussian distribution of thermal atoms, which is clearly identifiable from the wings of the density distribution. The temperature $T$ is related to the width $\sigma_i$ $\left(\text{where} \ i\in\left\{x,y,z\right\}\right)$ of the thermal fraction and the trap frequency $\omega_i$ via \cite{Wolfgang1999}
\begin{equation}
\label{temp_measurement}
T = \frac{m \sigma_i^2}{2 k_B}\frac{\omega_i^2}{1+\omega_i^2 t_\text{exp}^2}.
\end{equation}
As mentioned above, we measure the temperature using a single shot at an expansion time of $t_\text{exp}=24.1\,$ms. The temperature obtained this way is in agreement with a separate measurement of the thermal width $\sigma_i$ as a function of expansion time $t_\text{exp}$, using the temperature as the free parameter in a fit according to Eq. \ref{temp_measurement}.

\section{Observing quantum projection noise}
\label{results}

This section discusses the main results of the experiment in the context of the quantum projection noise limit. If other noise sources are sufficiently suppressed, the quantum projection noise is what determines the population fluctuations when measuring the state transfer with either a single $\frac{\pi}{2}$-pulse or a $\frac{\pi}{2}$-$\frac{\pi}{2}$-Ramsey sequence, potentially allowing us to detect spin-squeezed states in future experiments.

\subsection{Ramsey interferometry results}

\begin{figure}[t]
\includegraphics[scale=0.75]{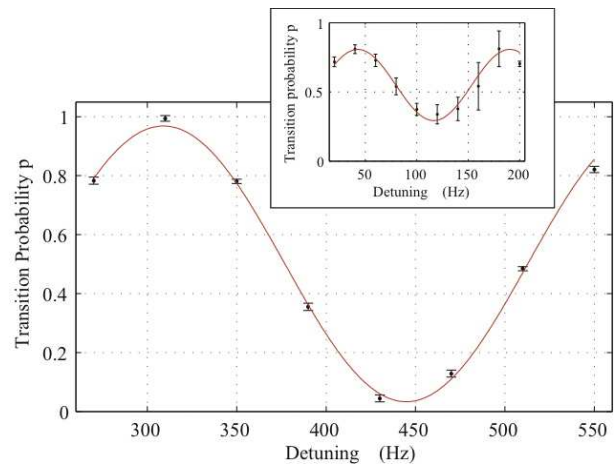}
\caption{\label{fringes_with_noise}(Color online) Noise measurement on a Ramsey fringe. The graph illustrates a significant improvement in signal-to-noise as compared to the data in \cite{D2009} (see inset).}
\end{figure}

As described in the previous section, each light sheet forms a $\frac{\pi}{2}$-pulse, and by changing the frequency driving the EOM, we scan through a sequence of Ramsey fringes. The central part of a typical fringe pattern is depicted in Fig. \ref{fig:fringes}, showing good visibility and a significantly improved signal-to-noise ratio as compared to our previous setup. Given the measured stability (see also Fig. \ref{fringes_with_noise} and Fig. \ref{noise_vs_number}), the best achievable phase sensitivity for a measurement averaged over five experimental runs ($300\,$s) is $1.3\,$mrad, an improvement of more than two orders of magnitude with respect to the stability in \cite{D2009}. 

The spectral widths of the beamsplitters are inversely proportional to the propagation time through each of the beamsplitters. It depends on the speed of the falling atom laser pulse and therefore on the exact position of the two light sheets with respect to the trapped BEC position. We have taken data for different configurations and do not quote a specific number for the spectral width. An order of magnitude estimate for the scales we are working with is $\simeq 1\,$kHz.

The significant improvement in the stability of the interferometer is mainly due to the new setup involving a Sagnac interferometer (see section \ref{Sagnac}). The passive stability in the system allows us to achieve standard deviations in the transition probability of below $0.5\,$\%, at the quantum projection noise limit for the atom numbers of $\simeq 10^4$ used here. Should a better stability of the Sagnac setup be required to achieve quantum projection noise limited operation at larger atom number, one could actively stabilize the interferometer and potentially achieve further improvement in performance. Figure \ref{fringes_with_noise} illustrates the improved signal-to-noise ratio presented in this work compared to the measurements using an actively stabilized Mach-Zehnder type setup. Each datapoint in the graph represents an average over ten measurements ($600\,$s measurement time per datapoint). The achieved stability of the Ramsey fringe signal is remarkable, given that the Raman setup relies on a fiber based Sagnac interferometer without active stabilization.

\subsection{Noise analysis}

\begin{figure}[t]
\includegraphics[scale=0.78]{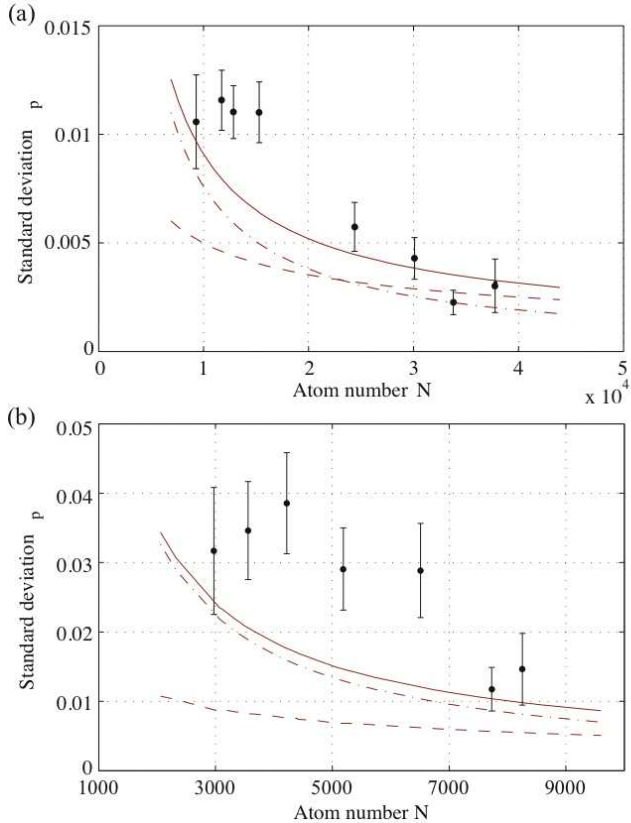}
\caption{\label{noise_vs_number}(Color online) Measurement of the standard deviation $\sigma_p$ as a function of total atom number N (a) for a single $\frac{\pi}{2}$-beamsplitter and (b) after a Ramsey $\frac{\pi}{2}$-$\frac{\pi}{2}$-sequence. The dashed line depicts the expected quantum projection noise, the dash-dotted line the photon shot noise, and the solid line the resulting standard deviation from both contributions. For sufficiently large atom numbers, the measured standard deviation agrees well with what is theoretically expected, showing that our apparatus operates at the quantum projection noise limit.}
\end{figure}

We observe noise in agreement with the quantum projection noise limit. The quantity that is analyzed is the transition probability $p=N_2/(N_1+N_2)$. The quantum projection noise limit is commonly defined in the context of (pseudo-)spin squeezing, where the pseudo-spin ${\bf S}$ describes the evolution of an effective two-level system \cite{W1993}. The third component of the pseudo-spin is given by $S_z=\frac{1}{2}(N_2-N_1)$, and the quantum projection noise limit on the variance of $S_z$ is $\sigma_{S_z}^2=N p (1-p)\simeq N/4$ for $p\simeq0.5$, the case of equal probability of the atoms being in state $\left|1\right\rangle$ and state $\left|2\right\rangle$. $N=N_1+N_2$ is the total number of atoms present in each run. The variance in the pseudo-spin projection translates directly into the variance in $p$, $\sigma_p^2=p (1-p)/N$. Due to the inherent normalization when calculating $p$, fluctuations in the total number $N$ do not significantly contribute to the variance $\sigma_p^2$, when $p\simeq0.5$. In any case, the calculated $\sigma_p$ will give a lower bound of the atomic noise present in the system.

An inherent source of noise in absorption imaging is the photon shot noise present in each single image taken with the CCD camera. The electrons collected in every pixel depend on the incident photon number via the camera's quantum efficiency, $N_\text{el}=QE\times N_\gamma$. The relevant variance to be considered is $\sigma_{N_\text{el}}^2=N_\text{el}$. Replacing $I$, $I_0$ and $I_\text{sat}$ in Eq. \ref{eq:imaging} with the corresponding electron counts makes it straightforward to determine the expected photon shot noise contribution to the atom number count:
\begin{equation}
\sigma_{\text{px},\gamma}^2 = c_0^2 \left[\left(\frac{L}{\sqrt{N_\text{el}}}+\frac{\sqrt{N_\text{el}}}{N_{\text{el,sat}}}\right)^2 + \left(\frac{L}{\sqrt{N_{\text{el},0}}}+\frac{\sqrt{N_{\text{el},0}}}{N_{\text{el,sat}}}\right)^2\right].
\end{equation}
Here, $N_\text{el}$, $N_{\text{el},0}$ and $N_\text{el,sat}$ are the pixelwise electron counts corresponding to the intensities $I$, $I_0$ and $I_\text{sat}$, respectively. From the electron count in each picture, we can accurately infer the noise contribution due to photon shot noise. For the transition probability $p$, the photon contribution scales as $\sigma_{p,\gamma}^2\propto 1/N^2$ (where $N$ is the total atom number), as compared to the scaling of $\sigma_p^2\propto1/N$ due to quantum projection noise. In the absence of additional noise sources and for sufficiently large atom numbers $N$, one expects the quantum projection noise to be the dominating source of uncertainty.

Our experimental images of the atom clouds are clean enough to be able to observe the quantum projection noise limit. For accurate atom number counting, we choose a relatively large region around each of the two atom laser pulse components, which have been separated by a Stern-Gerlach type experiment. A Thomas-Fermi distribution is then fitted to the integrated density profile, to determine the exact central position and width of each of the two clouds. Using this information, we choose a rectangular counting region to be centered around the fit center, with a width of $120\,\%$ of the fitted Thomas-Fermi diameter. Having measured each state's atom number independently, we calculate $p=N_2/(N_1+N_2)$. 

The quantum projection noise is strongest when choosing a point of equal numbers in states $\left|1\right\rangle$ and $\left|2\right\rangle$ ($p=0.5$), centered between a maximum and a minimum of the Ramsey fringes. We perform a stability measurement for $p=0.5$, both for a single $\frac{\pi}{2}$-beamsplitter and a Ramsey interferometer involving two $\frac{\pi}{2}$-pulses (Fig. \ref{noise_vs_number}). For different atom numbers $N$, we take sets of data at otherwise equal experimental conditions and measure the resulting variance $\sigma_p^2$ as a function of $N$. We observe standard deviations that for atom numbers $N\gtrsim 10^4$ are quantum projection noise limited. As the photon-caused variance $\sigma_{p,\gamma}$ scales as $1/N^2$, compared to the quantum projection noise scaling of $1/N$, it can be expected to dominate at low atom numbers whereas for larger atom numbers the quantum projection noise is dominant. Regarding their atom number dependence, other classical noise sources can be expected to behave similar to the photon shot noise. We see very good agreement of the experimentally measured noise with the theoretically expected standard deviation. For the data corresponding to a whole Ramsey interferometer, we were for technical reasons not able to measure at higher atom numbers; however, the data strongly suggests a behaviour similar to the one for the single beamsplitter.

\section{Conclusion}
\label{conclusion}

The data presented in this paper demonstrates the quantum projection noise limited performance of an atom laser interferometer. The results are of crucial importance for the realization of squeezing in free-space atom interferometry, in particular those involving coherent atoms extracted from a Bose-Einstein condensate. We measure uncertainties in agreement with the quantum projection noise limit, potentially allowing to observe an implementation of spin squeezing in future experiments. Spin squeezing will allow for interferometric sensitivity beyond the classical limit, yielding improved performance in high precision atom interferometry based measurement devices.

An atom laser like the one described here is well suited for the implementation of (high momentum) external state beamsplitting, and we intend to set up an external state Ramsey interferometer in the near future. We will use two counterpropagating beams containing only a single optical frequency each, with the splitting of the two frequencies given by the atomic ground state hyperfine splitting. Besides the fact that this has not been investigated before using a freely propagating atom laser, such a setup is interesting from the point of view of measuring spatial effects with the interferometer. In addition, it will allow for quadrature-squeezing in the atomic beam via nonlinear Kerr interactions \cite{Simon2009}.

\section{Acknowledgements}

We acknowledge financial support by the Australian Research Council Centre of Excellence program.

\end{document}